\documentstyle[preprint,aps,prd]{revtex}
\newcommand{\mlabel}[1]{\label{#1}}
\draft
\begin{document}
\title{
 \begin{flushright}{\normalsize HD--THEP--96--27    \\
                                 September 20, 1996 \\}
\end{flushright}
Classical Real Time Correlation Functions And Quantum Corrections
at Finite Temperature
}

\author{Dietrich B\"odeker
\footnote{e-mail: cq1@ix.urz.uni--heidelberg.de
}}

\address{ Institut f\"ur Theoretische Physik, Universit\"at Heidelberg,
Philosophenweg 16, 69120 Heidelberg, Germany}


\maketitle

\begin{abstract}
We consider quantum corrections to classical real time correlation
functions at finite temperature. We derive a semi-classical expansion
in powers of $\hbar$ with coefficients including all orders in the
coupling constant.  We give explicit expressions up to order
$\hbar^2$.  We restrict ourselves to a scalar theory. This method, if
extended to gauge theories, might be used to compute quantum
corrections to the high temperature baryon number violation rate in the
Standard Model.
\end{abstract}

\noindent

\narrowtext

\section{Introduction}

In the Standard Model baryon number is not conserved \cite{thooft}.
While the rate for baryon number violating processes is exponentially
suppressed at zero temperature it can be significant at temperatures
of the order of the electroweak scale
\cite{linde77,susskind,manton,kuzmin,arnold,khlebnikov}. This could
have had an influence on the baryon asymmetry of the universe (for a
recent review see \cite{rubakov}).

Baryon number nonconserving processes are mediated by topologically
nontrivial vacuum-to-vacuum transitions in the electroweak theory.
These transitions are characterized by a change $\Delta N_{\rm CS} =
\pm 1$  of the  the Chern--Simons number
\begin{eqnarray}
   N_{\rm CS} = \frac {g^2}{32\pi^2} \epsilon_{ijk} \int d^3x
  \left( F^a_{ij} A^a_k - \frac{g}{3} \epsilon^{abc} 
  A^a_i A^b_j A^c_k \right)
  \mlabel{chern}
\end{eqnarray}
where $A_{\mu}^a$ are the SU(2) gauge fields and $g$ is the weak
coupling constant. The change of baryon number $\Delta B$ is related
to $\Delta N_{\rm CS} $ by
\begin{eqnarray}
        \Delta B  = n_f \Delta N_{\rm CS} 
\end{eqnarray}
where $n_f$ is the number of fermion families.  At finite temperature
the $N_{\rm CS} $--changing transitions occur by thermal fluctuations
across an energy barrier.  For temperatures less than the critical
temperature of the electroweak phase transition $T_{\rm c}\sim 100$
GeV the transition rate can be calculated in a weak coupling coupling
expansion around the the so called sphaleron which is a saddle point
on this barrier.

The sphaleron approximation can not be extended to temperatures above
$T_{\rm c}$.  The field configurations which dominate the rate for
$T\gg T_{\rm c}$ have a typical size of order $\sim 1/(g^2 T)$
\cite{arnold}.  This is the scale at which perturbation theory at
finite temperature breaks down \cite{linde80}. For this temperature
range the baryon number violation rate per unit volume has been
estimated as \cite{arnold}
\begin{eqnarray}
  \Gamma = \kappa (\alpha_W T)^4
  \mlabel{Gamma}
\end{eqnarray}
where $\kappa$ is a dimensionless constant and $\alpha_W =
g^2/(4\pi)$. 

Grigoriev and Rubakov \cite{grigoriev} have suggested a numerical
algorithm for computing $\Gamma$. They argued that the thermal activation
across a barrier is a classically allowed process and that one should
determine the rate in a purely classical field theory.  This consists
of solving the classical equations of motion for the gauge fields and
measuring the change of the Chern--Simons number. Finally the average
over initial conditions has to be performed with the Boltzmann weight
$\exp(-\beta H)$. Since the relevant field modes are long wavelength
and involve many quanta one would expect that they in fact behave
classically.

The quantity of interest is the real time correlation function
\begin{eqnarray}
  C_{\rm CS} (t) = 
  \left\langle \left[N_{\rm CS}(t) - N_{\rm CS}(0)\right]^2 \right\rangle 
\end{eqnarray}
where $\langle \cdots \rangle $ denotes the average over a thermal
ensemble.  It is commonly assumed that the Chern--Simons number
performs a random walk so that for large times $C_{\rm CS}(t) $ 
grows linearly with $t$. Then the rate $\Gamma$ is related to 
$C_{\rm CS}(t) $ through
\begin{eqnarray}
   C_{\rm CS} (t) \to V \Gamma t \quad 
   \mlabel{random}
\end{eqnarray}
for $t \to \infty$ where $V$ is the space volume \cite{khlebnikov}.
Using real time evolution of classical equations of motion on a
Hamiltonian lattice Ambj\o rn et al.\ \cite{ambjorn91} obtained a
lower bound for $\Gamma$. Ambj{\o r}n and Krasnitz \cite{ambjorn95}
performed a more precise calculation for the pure SU(2) gauge theory,
they found $\Gamma = 1.1 (\alpha_W T)^4 $.  This result has been
confirmed by Moore \cite{moore96a,moore96b} who considered the change
of $N_{\rm CS}$ in the presence of a chemical potential for $N_{\rm
  CS}$. Tang and Smit \cite{smit} have used the algorithm of Ref.\ 
\cite{ambjorn95} to study the the SU(2)--Higgs model. For $T \gg
T_{\rm c}$ their results agree with those of Ref.\ \cite{ambjorn95}.
For temperatures less than $T_{\rm c}$ they found a result which is
much larger than the one obtained in the sphaleron approximation
\cite{kuzmin,arnold}. Moore and Turok \cite{moore96c} found similar
results for $T<T_{\rm c}$ but they argue that for this temperature
range the classical approxomation for the correlation function might
not be reliable.

There are several questions arising in this context. It has been known
since the work of Rayleigh, Jeans, Planck and Einstein that
statistical mechanics for a field theory is ill defined due to
ultraviolet divergences. Many physical quantities depend on the cutoff
and do not have a finite continuum limit when computed classically.
For static quantities the classical theory is equivalent to a three
dimensional Euclidean quantum field theory
\cite{ginsparg,appelquist,nadkarni,landsman,kajantie}. It is possible
to introduce counterterms in the classical Hamiltonian such that the
continuum limit exists. Since the three dimensional theory is
super--renormalizable only a finite number of divergences occur. For
time dependent quantities the situation is more complicated. In gauge
theories there are additional ultraviolet divergences not related to
the three dimensional theory \cite{bodeker}.  These arise in diagrams
which in the corresponding quantum theory give rise to terms $\propto
T^2$, the so called hard thermal loops \cite{pisarski}. These are
explicitly time dependent.  The hard thermal loops are due to large
loop momenta $k\sim T$ which are much larger than the momentum scale
$\sim g^2 T$ associated with the $N_{\rm CS}$--diffusion.  If the high
momentum modes decouple the continuum limit of certain quantities may
exist. The results of Refs.\ 
\cite{ambjorn95,moore96a,moore96b,smit,moore96c} indicate that this is
the case for $\Gamma$ in Eq.\ (\ref{random}) when computed for
temperatures much larger than $T_{\rm c}$.

If the continuum limit of the classical correlation function for some
particular quantity exists, the question remains how accurate the
classical approximation is. In Ref.\ \cite{moore96c} this problem has
been addressed perturbatively.  It would be desirable to find a way to
go beyond perturbation theory and to find an algorithm for incorporating
quantum corrections into the classical lattice calculations done so
far.

In the present paper we address the latter problem. We develop a
semi-classical expansion of real time correlation functions at finite
temperature starting from the quantum mechanical correlator. We keep
Planck's constant $\hbar\neq 1$. We derive an expansion in
powers of $\hbar$. In the limit $\hbar \to 0$ we recover the classical
limit. We consider correlation functions of the type
\begin{eqnarray}
  C(t) =  \left\langle \frac12 \Big[  A(0) B(t) + B(t) A(0) \Big]
  \right\rangle
  \mlabel{order}
\end{eqnarray}
where $A(t)$ and $B(t)$ are functions of the Heisenberg field
operators at time $t$. The operator ordering in Eq.\ (\ref{order}) is
the relevant one for the Chern--Simons number diffusion.
The problem of ultraviolet divergences related to real time evolution
will not be addressed here.

The semiclassical expansion for static finite temperature correlation
functions has been developed in the early days of quantum mechanics
\cite{wigner} and can be found in textbooks, e.g., Ref.\
\cite{landau}.  The author is not aware of such an expansion for time
dependent quantities existing in the literature.

In Section \ref{qm} we illustrate the method for the simplest quantum
mechanical system, i.e., for one degree of freedom.  The
generalization to a system with an arbitrary number of degrees of
freedom (Sect.\ \ref{many}) and thus to field theory
(Sect. \ref{fieldtheory}) is straightforward. Finally we summarize and
discuss the results in Sect.\ \ref{disc}.

\section{Quantum Mechanics with One Degree of Freedom}\mlabel{qm}

We consider one quantum mechanical degree of freedom $q$ with
conjugate momentum $p$ and the Hamiltonian $H= p^2/2 + U(q)$.  The
potential $U$ is assumed to be a fourth order polynomial in $q$.  The
quantum operators are denoted by capital letters.  Heisenberg and
Schr\"odinger operators are denoted by $Q(t)$, $P(t)$ and $Q=Q(0)$,
$P=P(0)$, respectively. We consider the two point function
\begin{eqnarray}
  C(t) = Z^{-1} 
  {\rm tr}\left\{ \frac12 \Big[Q(t)Q(0) + Q(0)Q(t)\Big]
  e^{-\beta H(P,Q)} \right\}
  \mlabel{c}
\end{eqnarray}
where $Z= {\rm tr}\left\{\exp(-\beta H)\right\}$. We will find an
expansion
\begin{eqnarray}
  C(t) = C_0(t) + \hbar C_1(t) + \hbar^2 C_2(t) + \cdots .
\end{eqnarray}

Following \cite{landau} we use (Schr\"odinger) momentum eigenstates
$|p\rangle$ to perform the trace and insert $1 = \int d q |q\rangle
\langle q|$ to obtain
\begin{eqnarray}
  \mlabel{ci}
  C(t) = Z^{-1} \int \frac{dp dq}{2\pi\hbar} I
\end{eqnarray}
with
\begin{eqnarray}
  \mlabel{i}
  I = \langle q|p\rangle  
  \langle p|\frac12 \Big[Q(t)Q(0) + Q(0)Q(t)\Big]
  e^{-\beta H(P,Q)}|q\rangle  .
\end{eqnarray}
Factorizing the Boltzmann factor one writes $I$ in terms of
$\chi$ as
\begin{eqnarray}
  I = \exp\{-\beta H(p,q)\} \chi
  \mlabel{chi}
\end{eqnarray}
Differentiating Eq.\ (\ref{i}) with respect to $\beta$ one obtains
\cite{landau}
\begin{eqnarray}
  \partial_{\beta} \chi&=& -i\hbar p \Big[ -\beta U'(q) \chi 
  + \partial_{q}\chi\Big]
  \nonumber \\ && {} + \frac12 \hbar^2 \Big[\partial_{q}^2 \chi
  -  2 \beta U'(q) \partial_{q}\chi
  - \beta U''(q)  \chi + \beta^2 U'(q)^2 \chi\Big] 
  \mlabel{dglchi}
\end{eqnarray}
where the primes denote derivatives with respect to $q$.  Eq.\
(\ref{dglchi}) can be solved by iteration when $\chi$ is expanded in
powers of $\hbar$,
\begin{eqnarray}
  \chi = \chi_0 + \hbar \chi_1 + \hbar^2 \chi_2 + \cdots
  \mlabel{expandchi}
\end{eqnarray}

We have to determine the boundary condition at $\beta = 0$,
\begin{eqnarray}
  \chi^0 &\equiv& \chi(\beta = 0)  = \langle q|p\rangle 
  \langle p|\frac12 (Q(t)Q(0) + Q(0)Q(t))|q\rangle   ,
\end{eqnarray}
which will be expanded in powers of $\hbar$ as well:
\begin{eqnarray}
  \chi^0 = \chi_0^0 + \hbar \chi_1^0 + \hbar^2 \chi_2^0 + \cdots .
  \mlabel{chi^0exp}
\end{eqnarray}
We write $\chi^0 $ as
\begin{eqnarray}
  \chi^0 = \left( q + \frac{i}{2}  \hbar \partial_p \right) \eta
  \mlabel{chi^0}
\end{eqnarray}
with 
\begin{eqnarray}
  \eta \equiv \langle q|p\rangle \langle p| Q(t) |q\rangle   .
\end{eqnarray}
Differentiating with respect to $t$ we find
\begin{eqnarray}
  \partial_t \eta &=& 
  \frac{i}{\hbar} \langle q|p\rangle 
  \langle p|[H(P,Q),Q(t)] |q\rangle 
  \mlabel{firstdgleta}
\end{eqnarray}
where we have used the fact that the Hamilton operator is
time--independent.  From Eq.\ (\ref{firstdgleta}) we obtain
\begin{eqnarray}
        \partial_t \eta &=& \frac{i}{\hbar} \Big( -i\hbar p
        + \frac{i}{2} \hbar^2 \partial_q^2 
        + U(q + i \hbar\partial_p ) -
        U(q) \Big) \eta \nonumber
\\
        &=& p \partial_q \eta - U'(q) \partial_p \eta 
        + \frac{i}{2}\hbar 
        \left(  \partial_q^2 \eta -  U''(q) \partial_p^2\eta  \right)
        +\frac16 \hbar^2  U'''(q) \partial_p^3 \eta 
        +\frac{i}{24}  \hbar^3  U^{(4)} \partial_p^4\eta .
        \mlabel{dgleta}
\end{eqnarray}
The initial condition for $\eta$  is 
\begin{eqnarray}
  \eta(t=0) = q .
  \mlabel{etaini}
\end{eqnarray}
We expand $\eta$ in powers of $\hbar$
\begin{eqnarray}
  \eta = \eta_0 + \hbar \eta_1 + \hbar^2 \eta_2  +\cdots
  \mlabel{1.8.3}
\end{eqnarray}
and solve Eq.\ (\ref{dgleta}) by iteration.
For $\eta_0$ we obtain the differential equation
\begin{eqnarray}
     \partial_t \eta_0 &=& \Big( p\partial_q - U'(q) \partial_p \Big)
     \eta_0 \nonumber
     \\
     &=& \{H, \eta_0 \}
     \mlabel{1.8.4}
\end{eqnarray}
where $\{,\}$ denotes the Poisson bracket
\begin{eqnarray}
  \{f,g\} = \partial_p f \partial_q g - \partial_p g .
  \partial_q f
\end{eqnarray}
The general solution to Eq. (\ref{1.8.4}) is $\eta_0 = f(p_{\rm
  c}(t),q_{\rm c}(t))$ where $p_{\rm c}(t)$ , $q_{\rm c}(t)$ are the
solutions to the classical Hamilton equations of motion
\begin{eqnarray}
     \dot{q}_{\rm c}(t) = \{ H,q_{\rm c}(t) \},   \quad 
     \dot{p}_{\rm c}(t) = \{ H,p_{\rm c}(t) \}
\end{eqnarray}
with the initial conditions
\begin{eqnarray}
  q_{\rm c}(0) = q, \quad p_{\rm c}(0) = p .
\end{eqnarray}
With Eq.\ (\ref{etaini}) we obtain
\begin{eqnarray}
  \eta_0 = q_{\rm c}(t)
\end{eqnarray}
and therefore $\chi_0^0 = q q_{\rm c}(t)$. From Eq.\ (\ref{dglchi}) we
see that $\chi_0$ is independent of $\beta$. Thus
\begin{eqnarray}
  \chi_0 = q q_{\rm c}(t) .
\end{eqnarray}
Then we finally obtain
\begin{eqnarray}
  C_0(t) = Z_0^{-1} \int \frac{dp dq}{2\pi\hbar}
  e^{-\beta H(p,q)}  q q_{\rm c}(t) 
  \mlabel{c0}
\end{eqnarray}
with $Z_0 = \int \frac{dp dq}{2\pi\hbar} e^{-\beta H(p,q)}$, which
corresponds to the Grigoriev-Rubakov expression.

Now we consider the order $\hbar$ correction to Eq.\ (\ref{c0}).
From Eq.\ (\ref{dglchi}) we obtain
\begin{eqnarray}
  \chi_1 = \chi_1^0 
  - i p \left( \beta \chi_0' - \frac12 \beta^2 U' \chi_0\right) .
  \mlabel{chi_1}
\end{eqnarray}
In the correlator $C(t)$ the function $\chi_1$ only appears in the
form $\int dp dq \exp(-\beta H) \chi_1$. One can integrate this
by parts to find the equivalent expression
\begin{eqnarray}
   \int dp dq e^{-\beta H(p,q)} \chi_1 
   &=& \int dp dq  e^{-\beta H(p,q)}
   \{ \chi_1^0 - \frac{i}{2} \partial_p \partial_q \chi_0 \} .
   \mlabel{chi_1equiv}
\end{eqnarray}
Eq.\ (\ref{chi^0}) gives
\begin{eqnarray}
  \chi_1^0 =  q \eta_1 + \frac{i}{2} \partial_p \eta_0
  \mlabel{chi_1^0}
\end{eqnarray}
where $\eta_1$ is determined by
\begin{eqnarray}
  \partial_t \eta_1 &=& \{H, \eta_1 \}
  + \frac{i}{2} \Big( \partial_q^2 - U''(q) \partial_p^2 \Big)
  \eta_0 
  \mlabel{1.8.6}
\end{eqnarray}
with the initial condition $\eta_1(t=0) = 0 $.  In order to solve Eq.\
(\ref{1.8.6}) it is convenient to rewrite it as
\begin{eqnarray}
  \partial_t \Big( \eta_1 
  - \frac{i}{2} \partial_p \partial_q \eta_0 \Big) 
  = \{H, \eta_1 
  - \frac{i}{2} \partial_p \partial_q \eta_0 \} .
\end{eqnarray}
Therefore
\begin{eqnarray}
  \eta_1 = \frac{i}{2} \partial_p \partial_q q_{\rm c}(t) .
  \mlabel{eta_1}
\end{eqnarray}
Combining Eqs.\ (\ref{chi_1^0}) and (\ref{eta_1}) we arrive at
\begin{eqnarray}
        \chi_1^0 = \frac{i}{2} \partial_p \partial_q \chi_0 .
        \mlabel{chi_1^0final} 
\end{eqnarray}
If now compare Eqs.\ (\ref{chi_1equiv}) and (\ref{chi_1^0final}) we
see that
\begin{eqnarray}
  C_1(t) = 0 ,
\end{eqnarray}
i.e.\ the order $\hbar$ correction to the classical correlation
function (\ref{c0}) vanishes.

Next we compute the ${ \cal O } ( \hbar^2 )$ corrections to $C_0$.
The $\beta$--dependent part of $\chi_2$ is determined by Eq.\
(\ref{dglchi}). Using the results for $\chi_1$ (Eqs.\ (\ref{chi_1})
and (\ref{chi_1^0final})) we obtain the lengthy expression
\begin{eqnarray}
        \chi_2 &=& \chi_2^0 + \frac12 \left(
        \beta \partial_q^2 \chi_0 - \beta^2 U'(q) \partial_q \chi_0
        - \frac12 \beta^2 U''(q) \chi_0
        + \frac13 \beta^3 ( U'(q) )^2 \chi_0 \right) \nonumber
\\
        && {} + p \left( \frac12 \beta \partial_p \partial_q^2 \chi_0
        - \frac14 \beta^2 U'(q) \partial_p \partial_q \chi_0 \right)
        \nonumber
\\
        && {} p^2 \left( -\frac12 \beta^2 \partial_q^2 \chi_0
        + \frac16 \beta^3 U''(q) \chi_0 
        + \frac12 \beta^3 U'(q) \partial_q \chi_0
        - \frac18 \beta^4 ( U'(q) )^2 \chi_0 \right) .
\end{eqnarray} 
It  can, however,  be simplified by partial integrations: 
\begin{eqnarray}
  \int dp dq e^{-\beta H(p,q)} \chi_2 &=& \int dp dq  e^{-\beta H(p,q)}
  \left\{ \chi_2^0
  +   \frac18 \partial_p^2 \partial_q^2 \chi_0 
    \right. \nonumber \\
  && \left. {}+ \frac{\beta}{24}  \left[ \partial_q^2 \chi_0
   + U''(q) \partial_p^2 \chi_0 - \beta U''(q) \chi_0 \right]\right\} .
\end{eqnarray}
We use Eq.\ (\ref{chi^0}) to write
\begin{eqnarray}
  \chi_2^0 =  q \eta_2 + \frac{i}{2} \partial_p \eta_1
  \mlabel{chi_2^0}
\end{eqnarray}
where $\eta_2$ satisfies 
\begin{eqnarray}
  \partial_t \eta_2 &=& \{H, \eta_2 \}
  + \frac{i}{2} \Big( \partial_q^2 - U''(q) \partial_p^2 \Big)
  \eta_1 + \frac{1}{6} U'''(q)\partial_p^3 \eta_0 \mlabel{1.8.7}
  \mlabel{dgleta_2}
\end{eqnarray}
with the initial condition $\eta_2(t=0)=0$.

To compute $\eta_2$ we rewrite Eq.\ (\ref{dgleta_2}) as
\begin{eqnarray}
  \partial_t \Big( \eta_2
  + \frac{1}{8} \partial_p^2 \partial_q^2 \eta_0 \Big) 
  = \{H, \eta_2
  + \frac{1}{8} \partial_p^2 \partial_q^2 \eta_0  \}
  + \frac{1}{24} U'''(q)\partial_p^3 \eta_0 .
\end{eqnarray}
Thus
\begin{eqnarray}
  \eta_2 = - \frac{1}{8} \partial_p^2 \partial_q^2 \eta_0 
  + \bar{\eta}_2
\end{eqnarray}
where  $\bar{\eta}_2$ satisfies
\begin{eqnarray}
  \partial_t \bar{\eta}_2 = \{H, \bar{\eta}_2\} 
  - \frac{1}{24} U'''(q) \{q,q_{\rm c}(t)\}_3 .
\end{eqnarray}
Here we have denoted 
\begin{eqnarray}
  \{f,g\}_0= g ,\qquad \{f,g\}_{n+1} = \{f,\{f,g\}_n\}
  \mlabel{poisson_n}
\end{eqnarray}
in order to write the derivatives $\partial_p$ in terms of Poisson
brackets. With the initial condition $\bar{\eta}_2(t=0)$ we find
\begin{eqnarray}
  \bar{\eta}_2 = - \frac{1}{24}
   \int_0^t dt'  U'''\big(q_{\rm c}(t')\big) 
   \{q_{\rm c}(t'),q_{\rm c}(t)\}_3 .
   \mlabel{eta_2bar}
\end{eqnarray}
Here we have used the fact that the Poisson bracket is invariant under
canonical transformations of the variables $p$ and $q$ and that the
time evolution can be viewed as a canonical transformation.

Putting everything together we can write the correlation function
(\ref{c}) as
\begin{eqnarray}
  C(t) &=& Z^{-1} \int \frac{dp dq}{2\pi\hbar}  e^{-\beta H(p,q)} 
  \left\{ \left[ 1  + \frac{\hbar^2\beta}{24} \left( \partial_q^2 
  +  U''(q) \partial_p^2 
  - \beta  U''(q) \right) \right] q q_{\rm c}(t)  + \hbar^2 q\bar{\eta}_2 
    \right\} \nonumber 
\\
   && {} + {\cal O} (\hbar^3)
\end{eqnarray}
where $\bar{\eta}_2 $ is given by Eq.\ (\ref{eta_2bar}).

\section{Many Degrees of Freedom}\mlabel{many}
The results obtained in the previous section can be easily generalized
to a system with an arbitrary number of bosonic degrees of freedom.  The
coordinates $q_i$ and momenta $p_i$ will be collectively denoted by
$q$ and $p$. The notation is completely analogous to the one in Sect.\
\ref{qm}. The Hamiltonian is $H= 1/2 \sum_i p_i^2 + U(q)$. We consider
the correlator
\begin{eqnarray}
        C(t) =  Z^{-1} 
         {\rm tr}\left\{\frac12 \Big[   A(Q) B(Q(t)) + B(Q(t)) A(Q) \Big]
        e^{-\beta H(P,Q)} \right\}
\end{eqnarray}
where $A(q)$ and $B(q)$ are some functions of $q$. We write $C(t)$ as
\begin{eqnarray}
        C(t) = Z^{-1} 
        \int \prod_i \left( \frac{dp_i dq_i}{2\pi \hbar} \right)
        e^{-\beta H(p,q)} \chi
\end{eqnarray}
The differential equation determining the $\beta$ dependence of
$\chi$ is \cite{landau}
\begin{eqnarray}
        \partial_{\beta} \chi &=& \sum_i \Bigg\{ -i \hbar p_i 
        \left( \partial_{q_i} \chi - \beta U_i(q) \chi \right)  
        \nonumber
\\
        && {} +\frac{\hbar^2}{2} \left( \partial_{q_i}^2 \chi 
        - 2 \beta  U_i(q) \partial_{q_i}\chi - \beta  U_{ii}(q) \chi
        + \beta^2 ( U_i(q) )^2 \chi \right) \Bigg\}
        \mlabel{chidglmany}
\end{eqnarray}
where a subscript $i$ on $U$ denotes the derivative $\partial_{q_i}$.

The boundary condition for Eq.\ (\ref{chidglmany}) is
$\chi(\beta=0) = \chi^0$ with 
\begin{eqnarray}
        \chi^0 = \frac12 \langle q|p\rangle 
        \langle p|[A(Q) B(Q(t)) + B(Q(t)) A(Q) ]|q\rangle  .
\end{eqnarray}
This can be written as
\begin{eqnarray}
         \chi^0 = \frac12 [ A(q) + A(q+i\hbar\partial_p) ] \eta
\end{eqnarray}
with 
\begin{eqnarray}
        \eta =  \langle q|p\rangle \langle p| B(Q(t)) |q\rangle  .
\end{eqnarray}
The time dependence of $\eta$ is determined by
\begin{eqnarray}
        \partial_t \eta &=& 
        \sum_i \left( p_i \partial_{q_i} \eta 
        - U_i(q) \partial_{p_i} \eta\right)
        + \frac{i}{2} \hbar \Big[ \sum_i \partial_{q_i}^2 \eta 
        -  \sum_{ij} U_{ij}(q) 
        \partial_{p_i}\partial_{p_j} \eta \Big]\nonumber
\\
        && {} + \frac16 \hbar^2 \sum_{ijk} U_{ijk}(q) 
        \partial_{p_i}\partial_{p_j} \partial_{p_k} \eta
        + \frac{i}{24} \hbar^3 \sum_{ijkl} U_{ijkl}(q) 
        \partial_{p_i}\partial_{p_j} \partial_{p_k} \partial_{p_l} \eta
        \mlabel{dgletamany} .
\end{eqnarray}
Note that that this differential equation is completely analogous to
Eq.\ (\ref{dgleta}). It does not involve the function $B$ which enters
only through the initial condition
\begin{eqnarray}
        \eta(t=0) = B(q) .
\end{eqnarray}
Again we expand $\eta = \eta_0 + \hbar\eta_1 + \hbar^2\eta_2 + \cdots$
and solve Eq.\ (\ref{dgletamany}) by iteration.  For the first term in
this expansion we obtain
\begin{eqnarray}
        \eta_0 = B(q_{\rm c}(t))
        \mlabel{eta0many}
\end{eqnarray}
where $q_{\rm c}(t) $ is the solution to the classical equations of
motion with initial condition $q_{\rm c}(0) =q $, $\dot{q}_{\rm c}(0)
=p $. The ${\cal O} (\hbar)$ correction to Eq.\ (\ref{eta0many}) is
given through
\begin{eqnarray}
        \eta_1 = \frac{i}{2} \sum_i \partial_{q_i}\partial_{p_i}\eta_0
\end{eqnarray}
and the result for $\eta_2$ is
\begin{eqnarray}
        \eta_2 = - \frac18 \sum_{ij} \partial_{q_i}\partial_{p_i}
        \partial_{q_j}\partial_{p_j}\eta_0 + \bar{\eta}_2
\end{eqnarray}
where
\begin{eqnarray}
         \bar{\eta}_2 = -\frac{1}{24} \int_0^t dt' U_{ijk}(q_{\rm c}(t'))
        \{ q_{{\rm c}i}(t'), \{ q_{{\rm c}j}(t'), \{ q_{{\rm c}k}(t'),
         B(q_{\rm c}(t))\}\}\} .
        \mlabel{eta_2barmany}
\end{eqnarray}

Now we can write down the first three terms of the expansion
(\ref{chi^0exp}) for $\chi^0$:
\begin{eqnarray}
        \chi_0^0 = A(q) B(q_{\rm c}(t))
\end{eqnarray}
\begin{eqnarray}
        \chi_1^0 = \frac{i}{2}\sum_i 
        \partial_{q_i}\partial_{p_i}\chi_0^0 
\end{eqnarray}
and
\begin{eqnarray}
        \chi_2^0 = -  \frac18 \sum_{ij}
        \left\{ \partial_{q_i}\partial_{p_i}
        \partial_{q_j}\partial_{p_j}\chi_0^0  
        + A_{ij}(q) \partial_{p_i} \partial_{p_j} B(q_{\rm c}(t)) \right\}
        + A(q) \bar{\eta}_2 
\end{eqnarray}
with $A_{ij}=\partial_{q_i}\partial_{q_j}A$.

Now that we know the boundary condition at $\beta = 0$ it is
straightforward to solve Eq.\ (\ref{chidglmany}) to determine $\chi$.
The result is somewhat lengthy but it simplifies after integrating
$\int \prod_i (d p_i dq_i )/(2\pi \hbar) \exp(-\beta H) \chi$ by
parts. As in Sect.\ \ref{qm} one finds that the ${\cal O}(\hbar)$
contribution to $C(t)$ vanishes and we obtain
\begin{eqnarray}
         C(t) &=& Z^{-1} \int \prod_i 
        \left( \frac{dp_i dq_i}{2\pi \hbar} \right)
        e^{-\beta H(p,q)} \nonumber
\\
  && \times \Bigg\{\Bigg[ 1  + \frac{\hbar^2\beta}{24} 
  \Bigg( \sum_i\partial_{q_i}^2
  +   \sum_{ij} U_{ij}(q) \partial_{p_i}\partial_{p_j}
  - \beta  \sum_i U_{ii}(q) \Bigg) \Bigg] A(q) B(q_{\rm c}(t)) \nonumber
\\
        && {} + \hbar^2 \Bigg( A(q) \bar{\eta}_2 
        - \frac18 \sum_{ij} 
        A_{ij}(q) \partial_{p_i}\partial_{p_j}B(q_{\rm c}(t))
        \Bigg)   \Bigg\} \nonumber 
\\
   && {} + {\cal O} (\hbar^3) .
        \mlabel{cfinalmany}
\end{eqnarray}

\section{Field Theory}\mlabel{fieldtheory}

The field theoretical expressions for the correlation function
are completely analogous to the ones found in the previous
section. We consider a theory with one scalar field $\varphi$
and its conjugate momentum $\pi$. 
We consider the correlation function 
\begin{eqnarray} 
  C(t )  = Z^{-1} 
  {\rm tr} \left\{ \frac12 
  \Big[ \Phi(0,\vec{x}_A) \Phi(t,\vec{x}_B) 
  + \Phi(t,\vec{x}_B) \Phi(0,\vec{x}_A) \Big]
  e^{-\beta H[\Pi,\Phi]} \right\}
  \mlabel{cfield}
\end{eqnarray}
for the Hamiltonian 
\begin{eqnarray}
        H[\pi,\varphi] = \int d^3x \frac12 \pi^2(\vec{x}) + U[\varphi]
\mlabel{hfield}
\end{eqnarray}
where
\begin{eqnarray}
        U[\varphi] = \int d^3x \left\{
        \frac12 (\nabla \varphi(\vec{x}))^2 + V(\varphi(\vec{x})) 
        \right\} . \mlabel{U}
\end{eqnarray}
Replacing 
\begin{eqnarray}
        \sum_i \partial_{p_i}\partial_{q_i} \to 
        \int d^3 x \frac{\delta}{\delta \pi(\vec{x})}
        \frac{\delta}{\delta \varphi(\vec{x})}
\end{eqnarray}
the $\hbar$ expansion of (\ref{cfield}) can be read off from Eqs.\
(\ref{eta_2barmany}) and (\ref{cfinalmany}):
\begin{eqnarray}
 C(t) &=& Z^{-1} \int [d\pi d\varphi]
  e^{-\beta H[\pi,\varphi]} \Bigg\{ \Bigg[ 1 + 
  \frac{\hbar^2\beta}{24} \int d^3 x \Bigg( 
  \frac{\delta^2}{\delta \varphi(\vec{x})^2 } \nonumber
\\
  && {} + \int d^3 x' \delta(\vec{x} - \vec{x}') 
  \left( - \nabla^2_x  + V''(\varphi(\vec{x}) \right) 
  \frac{\delta^2 }{ \delta \pi(\vec{x}) \delta \pi(\vec{x}') }
  \nonumber 
\\
   && {} 
  - \beta  \frac{\delta^2 U[\varphi]}{\delta \varphi(\vec{x})^2 } 
  \Bigg)
  \Bigg]   \varphi(\vec{x}_A) \varphi_{\rm c}(t,\vec{x}_B)
   + \hbar^2 \varphi(\vec{x}_A) \bar{\eta}_2 \Bigg\} +
   {\cal O}(\hbar^3) 
  \mlabel{cfinalfield}
\end{eqnarray}
where $ \varphi_{\rm c}$ is the classical field satisfying
the equation of motion 
\begin{eqnarray}
  \ddot{\varphi_{\rm c}} - \nabla^2 \varphi_{\rm c} 
  + V'( \varphi_{\rm c}) = 0
\end{eqnarray}
with the initial conditions 
\begin{eqnarray}
  \varphi_{\rm c}(0,\vec{x}) = \varphi(\vec{x}), \quad
  \dot{\varphi}_{\rm c}(0,\vec{x}) = \pi(\vec{x}) .
\end{eqnarray}
Furthermore, $\bar{\eta}_2$ in Eq.\ (\ref{cfinalfield}) is given by
\begin{eqnarray}
  \bar{\eta}_2 = -\frac{1}{24} \int_0^t dt' \int d^3 x
  V'''(\varphi_{\rm c}(t',\vec{x}))
  \{\varphi_{\rm c}(t',\vec{x}), \varphi_{\rm c}(t,\vec{x}_B) \}_3
  \mlabel{eta_2barfield}
\end{eqnarray}
where $\{,\}_3$ is defined as in Eq.\ (\ref{poisson_n}) and the
Poisson brackets are
\begin{eqnarray}
  \{f,g\} = \int d^3x \left(\frac{\delta f}{\delta \pi(\vec{x})}
  \frac{\delta g}{\delta \varphi(\vec{x})} -
  \frac{\delta g}{\delta \pi(\vec{x})}
  \frac{\delta f}{\delta \varphi(\vec{x})} \right) .
\end{eqnarray}

The right hand side of Eq.\ (\ref{cfinalfield}) contains ultraviolet
divergences. Consider, e.g., the term $\delta^2 U[\varphi]/\delta
\varphi(\vec{x})^2 $. The singularities arise because we take two
functional derivatives of $U$ with respect to $\varphi(\vec{x})$, both
of them for the same space point $\vec{x}$.  If we write $\delta^2
U[\varphi]/\delta \varphi(\vec{x})^2 $ in momentum space and introduce
a momentum cutoff $\Lambda$ it becomes
\begin{eqnarray}
        \frac{\delta^2 U[\varphi]}{\delta \varphi(\vec{x})^2 } =
        \frac{1}{2\pi^2}
        \Bigg(\frac15 \Lambda^5 + \frac13 \Lambda^3
        V''(\varphi(\vec{x}) ) \Bigg) .
\end{eqnarray}

The parameters in the Hamiltonian (\ref{hfield}) are
bare quantities which have to be expressed in terms of renormalized
ones. In the following we will see how the singularities are canceled
after renormalization in the limit $t\to 0$ of $C(t)$. In this limit
the problem is of course much simpler. Ultraviolet divergences
related to real time evolution have been discussed in Ref.\
\cite{bodeker} and some of them turned out to be rather nontrivial
even in the classical limit i.e.\ without including the ${\cal O}
(\hbar^2)$ corrections.  As it was already mentioned the numerical
calculations done so far indicate that nevertheless the continuum
limit of the classical ${\cal O} (\hbar^0)$ result for $\Gamma$ (Eq.\
(\ref{random})) exists. Therefore it might even be meaningful to
consider quantum corrections leaving aside ultraviolet problems for a
moment.

For definiteness we choose the potential in (\ref{U}) as
\begin{eqnarray}
        V(\varphi) = \frac{m^2}{2} \varphi^2 + 
        {\lambda \over {4!}} \varphi^4 .
\end{eqnarray}
For $t = 0$ the path integral over $\pi$ is Gaussian and $\pi$ can be
integrated out. This  gives a constant factor which is canceled by the
corresponding factor in $Z$ and which will be
ignored in the following. Eq.\ (\ref{cfinalfield}) becomes
\begin{eqnarray}
  C(0) &=& Z^{-1} \int [d\varphi] e^{-\beta U[\varphi]}
   \nonumber 
\\
  && \times \Bigg\{ \Bigg[ 1  -
  \frac{\hbar^2\beta^2}{24} \frac{1}{2\pi^2}\int d^3 x 
  \Bigg(\frac15 \Lambda^5 + \frac13 \Lambda^3
  \Big(m^2 + \frac{\lambda}{2} \varphi^2(\vec{x}) \Big)\Bigg)
  \Bigg]   \varphi(\vec{x}_A) \varphi(\vec{x}_B) \nonumber
\\
  &&{} + \frac{\hbar^2\beta}{12} \delta(\vec{x}_A - \vec{x}_B)
  \Bigg\} +  {\cal O}(\hbar^3) 
  \mlabel{cstaticregular} .
\end{eqnarray}

We have to express the bare parameters in terms of the renormalized
ones which we denote by a subscript r. We write 
\begin{eqnarray}
        U = U_{\rm r} + \delta U .
\end{eqnarray}
We include a field independent part in the Hamiltonian which we denote
by $U^0$. The counterterm $\delta U^0$ is written as a power series in
$\hbar$, $\delta U^0 = \delta U^0_0 + \hbar^2 \delta U^0_2 + \cdots $.
Then one has to choose
\begin{eqnarray}
        \delta U^0_2 = - \frac{\beta}{24} \frac{1}{2\pi^2}\int d^3 x 
        \Bigg(\frac15 \Lambda^5 + \frac13 \Lambda^3 m^2 \Bigg)
\end{eqnarray}
to cancel the first two singular terms in Eq.\ (\ref{cstaticregular}).

For the mass renormalization we write
\begin{eqnarray}
         m_{\rm r}^2 = m^2 - \delta m^2 .
\end{eqnarray}
We define the renormalized mass in the imaginary
time formalism (see, e.g., \cite{kapusta}):
\begin{eqnarray}
        m^2_{\rm r} = m^2 + \Pi(\omega_n = 0, \mu^2 )
        \label{rencond}
\end{eqnarray} 
where $\Pi(\omega_n , \vec{k}^2 )$ is the self energy for Matsubara
frequency $\omega_n = n 2\pi T/\hbar$ and momentum $\vec{k}$ and $\mu$
is some renormalization point. If $\mu$ is chosen sufficiently large,
$m^2_{\rm r} $ can be calculated in perturbation theory.  The ${\cal
  O}(\lambda) $ contribution to $ \delta m^2 $ is given by the diagram
(a) in Fig.\ref{oneloopmass}:
\begin{eqnarray}
  \delta m^2 = - \frac{\lambda}{2} T \sum_{n=-\infty}^{\infty}
   \int\frac{d^3 k}{(2\pi)^3} \frac{1}{ \omega_n^2 + \vec{k}^2 + m^2 }
  + {\cal O} (\lambda^2) .
  \mlabel{deltam}
\end{eqnarray}
We can expand $ \delta m^2 $ in powers of $\hbar$,
\begin{eqnarray}
  \delta m^2 = \delta m^2_0 + \hbar^2 \delta m^2_2 + {\cal O}(\hbar^4) .
\end{eqnarray}
The zero Matsubara frequency part of Eq.\ (\ref{deltam}) gives 
\begin{eqnarray}
   \delta m^2_0 =  - \frac{\lambda_{\rm r}}{2} \int\frac{d^3 k}{(2\pi)^3} 
   \frac{1}{ \vec{k}^2 + m^2 } + {\cal O} (\lambda^2) .
\end{eqnarray}
The nonzero Matsubara frequency part can be expanded in powers of
$\hbar^2$. With a momentum cutoff $\Lambda$ the first term in
this series gives
\begin{eqnarray}
   \delta m^2_2 =  
   - \frac{\lambda\beta}{24} \frac{1}{2\pi^2} \frac{\Lambda^3}{3}
   \mlabel{deltam2} .
\end{eqnarray}
The zero Matsubara frequency parts of all higher loop diagrams for
$\delta m^2 $ (e.g. the diagram (b) in Fig.\ \ref{oneloopmass})
contribute to $\delta m_0^2 $. However if one goes beyond ${\cal
O}(\hbar^0)$ there are always at least two propagators in which one
particular nonzero Matsubara frequency appears. Therefore the $\hbar$
expansion for these diagrams beyond ${\cal O}(\hbar^0)$ starts at
${\cal O}(\hbar^4)$ and Eq.\ (\ref{deltam2}) is valid to all orders in
$\lambda$. For the coupling constant and wave function renormalization
there is no ${\cal O}(\hbar^2)$ contribution even at the one loop
level. In general there are finite counterterms of the order $\hbar^0$
which, however, do not have to be specified for the present purpose.

Replacing $U\to U_{\rm r} + \delta U_0 + \hbar^2 \delta U_2 $ in Eq.\
(\ref{cstaticregular}) and expanding in powers of $\hbar$ the ${\cal
O}(\hbar^2)$ divergences in Eq.\ (\ref{cstaticregular}) cancel and we
obtain
\begin{eqnarray}
 C(0) &=& Z^{-1} \int [d\varphi] 
  \exp\left\{-\beta (U_{\rm r}[\varphi] + \delta U_0[\varphi])\right\} 
  \nonumber 
\\
  && \times \Bigg\{  \varphi(\vec{x}_A) \varphi(\vec{x}_B) 
   + \frac{\hbar^2\beta}{12} \delta(\vec{x}_A - \vec{x}_B)
  \Bigg\} +  {\cal O}(\hbar^3) 
  \mlabel{cstaticfinite}
\end{eqnarray}
This expression is similar to a dimensionally reduced theory
\cite{kajantie}. There are two differences. First, the renormalized
parameters in Eq.\ (\ref{cstaticfinite}) are different from the ones
in Ref.\ \cite{kajantie}. This difference is due to finite parts in
the counterterm $\delta U_0[\varphi]$ corresponding to different
renormalization conditions.  The choice in Ref.\ \cite{kajantie} was
made such that the Greens functions in the 3-dimensional theory and
the Greens functions with $\omega_n =0$ in the 4-dimensional theories
agree up to a certain order in the coupling constant. Furthermore, in
Ref.\ \cite{kajantie} the renormalized mass was written in terms of
the physical Higgs mass.

The second difference between (\ref{cstaticfinite}) and the
expressions given in Ref.\ \cite{kajantie} is the appearance of the
$\delta$--function in Eq.\ (\ref{cstaticfinite}).  This term arises
because in the formalism described in the present paper there is no
separation of the fields into zero and nonzero Matsubara frequency
components. Therefore, the correlation function (\ref{cstaticfinite})
also contains the diagrams in which the external lines have nonzero
Matsubara frequency. In momentum space the tree level propagator for
the $\omega_n \neq 0$ fields is $D(\omega_n, \vec{k}^2) = 1/(
\omega_n^2 + \vec{k}^2)$. Expanding this expression in powers of
$\hbar$ gives $ \hbar^2/(2\pi n T)^2 + {\cal O} (\hbar^4)$. Summing
over all $n\neq  0$ we obtain the $\delta$--function--term in Eq.\
(\ref{cstaticfinite}).  All higher loop contributions to $D(\omega_n,
\vec{k}^2) $ are $ {\cal O} (\hbar^4)$ or higher.  Therefore they do
not contribute to $C(0)$ at the order considered here.

\section{Summary and Discussion}\mlabel{disc}
For real time correlation functions at finite temperature we have
obtained an expansion in powers of Planck's constant $\hbar$ which, at
lowest order, corresponds to the classical correlation function in the
sense of Grigoriev and Rubakov \cite{grigoriev}.  We have restricted
ourselves to a scalar theory. For the correlation functions of the
type (\ref{order}) we found that the first order correction in $\hbar$
vanishes. The first nonvanishing contribution is of the order
$\hbar^2$. The result has been written in terms of the solution to the
classical field equations and it's functional derivatives with respect
to the initial conditions. The expansion outlined in this paper can be
easily extended to higher orders in $\hbar$.

The method outlined in this paper might serve as a starting point
for numerical computations of quantum corrections to 
classical correlation functions which can not be calculated
in perturbation theory. An important example is the Chern-Simons
number diffusion rate in the electroweak theory which determines
the rate for baryon number violating processes in the early 
universe. 

The ${\cal O} (\hbar^2)$ correction to the classical correlation
function contains a term (\ref{eta_2barfield}) which is nonlocal in
time. One may ask how this term can be implemented in a numerical
evaluation of the correlation function.  It might be possible to find
a set of local equations of motion such that their solution reproduces
Eq.\ (\ref{eta_2barfield}). In Eq.\ (\ref{cfinalfield}) the
derivatives with respect to the initial conditions can be transformed
into local operators at $t = 0$ through partial integrations of the
path integral.

An important task is to extend the method outlined here to gauge
theories.  Another important issue are the ultraviolet divergences
of the classical result and of the ${\cal O} (\hbar^2)$
corrections. For the pure gauge theory considered in Refs.\
\cite{ambjorn95,moore96a} this maybe not so urgent since there the
classical results seem to have a well behaved continuum limit. Then
the ultraviolet behavior of the ${\cal O} (\hbar^2)$ corrections
may not be that  critical either. However, if the Higgs field is
included as in Ref.\ \cite{smit} the problem of ultraviolet
divergences is crucial.




\section*{Acknowledgments}

I am grateful to W.\ Buchm\"uller, A.\ Kovner, M.\ Laine, L.\ McLerran
and M.\ G.\ Schmidt for helpful and stimulating discussions.

\begin{figure}
\begin{picture}(200,100)
  \put(20,50){\line(1,0){25} }
  \put(70,50){\line(-1,0){25} }
  \put(70,70){\circle{40} }
  \put(70,10){(a)}
  \put(70,50){\line(1,0){50} }
\end{picture}
\begin{picture}(200,100)
  \put(30,50){\line(1,0){20} }
  \put(70,50){\line(-1,0){20} }
  \put(90,50){\circle{40} }
  \put(90,10){(b)}
  \put(90,30){\line(-1,0){1} }
  \put(70,50){\line(1,0){80} }
\end{picture}
\caption{Feynman diagrams contributing to mass renormalization.}
\mlabel{oneloopmass}
\end{figure}


\begin{thebibliography} {99}

\bibitem{thooft}
G.\ t'Hooft, Phys.\ Rev.\ Lett.\ {\bf 37} (1976) 8;
Phys.\ Rev.\ D {\bf 14} (1976) 3432.

\bibitem{linde77}
A.\ D.\ Linde, Phys.\ Lett.\ {\bf 70B} (1977) 306.
 
\bibitem{susskind}
S.\ Dimopoulos, L.\ Susskind, Phys.\ Rev.\ D {\bf 18} (1978) 4500.

\bibitem{manton}
N.\ Manton, Phys.\ Rev.\ D {\bf 28} (1983) 2019;
F.\ Klinkhamer, N.\ Manton,
Phys.\ Rev.\ D {\bf 30} (1984) 2212.

\bibitem{kuzmin} V.\ A.\ Kuzmin, V.A.\ Rubakov, M.\ E.\ Shaposhnikov,
Phys.\ Lett.\ {\bf 155B} (1985) 36.

\bibitem{arnold} P.\ Arnold, L.\ McLerran, Phys.\ Rev.\  D {\bf
36} (1987) 581.
 
\bibitem{khlebnikov} S.\ Yu.\ Khlebnikov, M.\ E.\ Shaposhnikov,
Nucl.\ Phys.\ {\bf B308} (1988) 885.

\bibitem{rubakov} V.A.\ Rubakov, M.\ E.\ Shaposhnikov, 
Usp.\ Fis.\ Nauk {\bf 166} (1996)  493,
hep-ph/9603208.

\bibitem{linde80} A.\ D.\ Linde, Phys.\ Lett.\ {\bf 96B (1980)} 289.

\bibitem{grigoriev}
D.Yu. Grigoriev and V.A.\ Rubakov, Nucl.Phys. {\bf B299} (1988) 67.

\bibitem{ambjorn91}
J.\ Ambj\o rn et al., Nucl.\ Phys.\  {\bf B253}  (1991) 346; 
J.\ Ambj\o rn and K. Farakos, Phys.\ Lett.\  
{\bf B294} (1992) 248.

\bibitem{ambjorn95}
J.\ Ambj\o rn, A. Krasnitz, Phys.\ Lett.\  {\bf B362} (1995) 97.

\bibitem{moore96a} G.\ D.\ Moore, hep-ph/9603384.

\bibitem{moore96b} G.\ D.\ Moore, hep-lat/9605001.

\bibitem{smit} W.\ H.\ Tang, J.\ Smit, hep-lat/9605016.

\bibitem{moore96c} G.\ D.\ Moore, N.\ Turok, hep-ph/9608350.

\bibitem{ginsparg} P.\ Ginsparg, Nucl.\ Phys.\ {\bf B170} (1980) 388.

\bibitem{appelquist} T.\ Appelquist, R.\ Pisarski, Phys.\ Rev.\ D
{\bf 23} (1981) 2305.

\bibitem{nadkarni} S.\ Nadkarni, Phys.\ Rev.\ D {\bf 27} (1983) 917.

\bibitem{landsman} N.P.\ Landsman, Nucl.\ Phys.\ {\bf B322} (1989)
498.

\bibitem{kajantie} K.\ Kajantie, M.\ Laine, K.\ Rummukainen,
M.\ Shaposhnikov, Nucl.\ Phys.\ {\bf B458} (1996) 90.

\bibitem{bodeker} D.\ B\"odeker, L.\ McLerran, A.\ Smilga, 
Phys.\ Rev.\ D {\bf 52} (1995) 4675.

\bibitem{pisarski}
E. Braaten and R. Pisarski, Nucl. Phys. {\bf B337} (1990) 569; 
Phys. Rev. D
{\bf 45} (1992) 1827; 
J. Frenkel and J.C. Taylor, Nucl. Phys. {\bf B334} (1990)  199; 
J.C. Taylor and S.M.H. Wong, Nucl.\ Phys.\ {\bf B346} (1990) 115;
R.\ Efraty, V.P. Nair, Phys.Rev.D {\bf 47} (1993) 5601; 
J.P.\ Blaizot, E.\ Iancu, Nucl.Phys.\ {\bf B417} (1994) 608; 
V.P. Nair, Phys.Rev.D {\bf 50} (1994) 4201.  

\bibitem{wigner} E.\ Wigner, Phys.\ Rev.\ {\bf 40} (1932) 749;
G.\ E.\ Uhlenbeck, L.\ Gropper, Phys.\ Rev.\ {\bf 41} (1932) 79;
J.\ G.\ Kirkwood, Phys.\ Rev.\ {\bf 44} (1933) 31.

\bibitem{landau}
E.\ M.\  Lifshitz and L.P.\ Pitaevsky,  Statistical Physics
(Pergamon Press, Oxford, 1981).

\bibitem{kapusta} J.\ I.\ Kapusta, Finite--temperature field theory
  (Cambridge University Press, 1989).

\end{thebibliography}
\end{document}